# Protection of Ising spin-orbit coupling in bulk misfit superconductors

Tomas Samuely[1,✉], Darshana Wickramaratne[2], Martin Gmitra[1,3], Thomas Jaouen[4], Ondrej Šofranko[1], Dominik Volavka[1], Marek Kuzmiak[3], Jozef Haniš[1], Pavol Szabó[3], Claude Monney[5], Geoffroy Kremer[5,6], Patrick Le Fèvre[7], François Bertran[7], Tristan Cren[8], Shunsuke Sasaki[9], Laurent Cario[9], Matteo Calandra[10], Igor I. Mazin[11,✉], Peter Samuely[3]

[1]Centre of Low Temperature Physics, Faculty of Science, Pavol Jozef Šafárik University in Košice, 04001 Košice, Slovakia

[2]Center for Computational Materials Science, U.S. Naval Research Laboratory, Washington, DC 20375, USA

[3]Centre of Low Temperature Physics, Institute of Experimental Physics, Slovak Academy of Sciences, 04001 Košice, Slovakia

[4]Univ Rennes, CNRS, (IPR Institut de Physique de Rennes) - UMR 6251, F-35000 Rennes, France

[5]Département de Physique and Fribourg Center for Nanomaterials, Université de Fribourg, Fribourg CH-1700, Switzerland

[6]Institut Jean Lamour, UMR 7198, CNRS-Université de Lorraine, Campus ARTEM, 2 allée André Guinier, BP 50840, 54011 Nancy, France

[7]SOLEIL synchrotron, L'Orme des Merisiers, Départementale 128, F-91190 Saint-Aubin, France

[8]Institut des NanoSciences de Paris, Sorbonne Université and CNRS-UMR 7588, Paris 75005, France

[9]Institut des Matériaux Jean Rouxel, Université de Nantes and CNRS-UMR 6502, Nantes 44322, France

[10]Dipartimento di Fisica, Università di Trento, via Sommarive 14, I-38123 Povo, Italy

[11]Department of Physics and Astronomy and Quantum Science and Engineering Center, George Mason University, Fairfax, VA 22030, USA

✉e-mail: tomas.samuely@upjs.sk, imazin2@gmu.edu

**Low-dimensional materials have remarkable properties that are distinct from their bulk counterparts. A paradigmatic example is Ising superconductivity that occurs in monolayer materials such as NbSe$_2$ which show a strong violation of the Pauli limit. In monolayers, this occurs due to a combination of broken inversion symmetry and spin-orbit coupling that locks the spins of the electrons out-of-plane. Bulk NbSe$_2$ is centrosymmetric and is therefore not an Ising superconductor. We show that bulk misfit compound superconductors, (LaSe)$_{1.14}$(NbSe$_2$) and (LaSe)$_{1.14}$(NbSe$_2$)$_2$, comprised of monolayers and bilayers of NbSe$_2$, exhibit unexpected Ising protection with a Pauli-limit violation**

**comparable to monolayer NbSe$_2$, despite formally having inversion symmetry. We study these misfit compounds using complementary experimental methods in combination with first-principles calculations. We propose theoretical mechanisms of how the Ising protection can survive in bulk materials. We show how some of these mechanisms operate in these bulk compounds due to a concerted effect of charge-transfer, defects, reduction of interlayer hopping, and stacking. This highlights how Ising superconductivity can, unexpectedly, arise in bulk materials, and possibly enable the design of bulk superconductors that are resilient to magnetic fields.**

In superconductors, an external magnetic field can break the Cooper pairs via two mechanisms. The first is orbital pair breaking due to the Lorentz force exerted by this field. This does not occur in two-dimensional (2D) superconductors if the magnetic field is in-plane. The second mechanism is spin pair breaking due to Zeeman shifts of opposite signs on the antiparallel spins in a spin-singlet Cooper pair. The minimum field to break pairs is the Pauli paramagnetic limit $B_P$[1,2]. It can be understood from the energy balance: in the normal (but not in the superconducting) state, an external field $B$ is Pauli-screened, with an energy gain of $\chi_{\text{Pauli}}B^2/2$, where $\chi_{\text{Pauli}}$ is the Pauli susceptibility determined by the density of states, $N(E_F)$, at the Fermi level ($E_F$). In the Bardeen-Cooper-Schrieffer (BCS) theory the pairing energy is $N(E_F)\Delta^2/2$, where $\Delta$ is the superconducting energy gap. Hence, when $B_P$ exceeds $\Delta$, it is energetically favourable to destroy superconductivity.

A superconducting state that can screen an external field is not subject to Pauli-limiting pair breaking, such as in some triplet superconductors[3]. Recently discovered Ising superconductors (IS) are also not Pauli limited for in-plane magnetic fields. Their spin-orbit coupling (SOC) splits the bands and leads to electron spins that are perpendicular to the 2D plane with the spins of the SOC-partner states aligned antiparallel. One consequence is that the Pauli screening for in-plane magnetic fields is effected by the SOC split states that are removed from the Fermi level and therefore not affected by superconductivity[4]. As long as both the Zeeman shift and $\Delta$ are small compared to the SOC splitting (i.e., many hundreds of Tesla in NbSe$_2$) the Pauli pair breaking is absent[5–12]. This is not unrelated to the resilience of triplet superconductivity to magnetic fields: in IS half of the Cooper pairs are described as singlet + triplet, and half as singlet - triplet[4]. The absence of inversion symmetry is a prerequisite for SOC to lift the Kramers' degeneracy. In the canonical IS, NbSe$_2$, this is satisfied in a monolayer, but not in a bilayer or bulk, preventing IS in bulk NbSe$_2$[5]. In this study we report experiments and theory that clearly demonstrate that bulk layered compounds containing NbSe$_2$ can exhibit Ising protection[13,14].

## Results

Figure 1a, b illustrates the two compounds that we synthesize and investigate, $(LaSe)_{1.14}(NbSe_2)$ and $(LaSe)_{1.14}(NbSe_2)_2$, which we refer to as 1T1H and 1T2H, respectively. Previous work has shown that both are superconducting with $T_c$ = 1.23 K and 5.7 K, respectively, and they both exhibit an in-plane critical field $B_{c2//ab}$ that strongly exceeds the Pauli limit $B_P$[13]. For $(LaSe)_{1.14}(NbSe_2)_2$ the ratio of $B_{c2//ab}/B_P$ is ~ 5, for $(LaSe)_{1.14}(NbSe_2)$ it is ~10 – a clear signature of IS in bulk compounds. Their fundamental structural units are monolayers of $NbSe_2$, where Nb atoms are in a trigonal prismatic coordination with Se, and LaSe with a rock salt structure. A stoichiometric compound of $NbSe_2$ and LaSe forms when one of the constituent substructures distorts to accommodate the lattice parameters of the second substructure, hence, they are often referred to as misfit compounds[15]. One unit cell of $NbSe_2$ has a basal plane area of 10.25 Å² [16], while one unit cell of LaSe has an area of 9.19 Å² [17]; their ratio is larger than 1 and leads to the two misfit compounds we investigate.

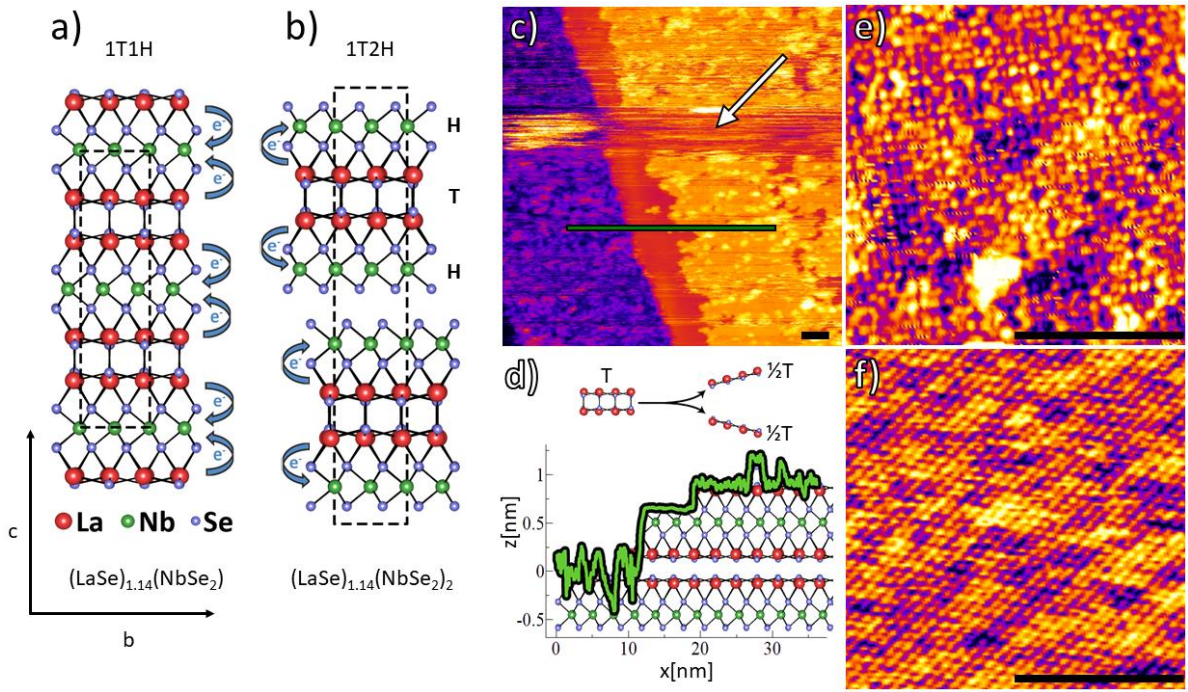

**Figure 1 – Bulk misfit compounds that exhibit Ising superconductivity.** Crystal structure of (a) 1T1H and (b) 1T2H. Dashed rectangles represent the unit cell in the *bc* plane. c) Topography of three atomic terraces of 1T1H, H surface in the middle and ½T surfaces adjacent to the H surface. The white arrow indicates the area, where the scanning tip disrupted the unstable ½T layer. d) (Top) Schematic illustration of how the T layer of 1T1H can be cleaved into two ½T layers. (Bottom) Height profile along the green line in (c), imposed on the illustrated atomic structure of the H and T layers. The vertical size of the illustrated atomic structure is to scale while the horizontal size is exaggerated for clarity. The height profile reveals the step size between the NbSe$_2$, and ½T layer terrace is 3 Å, i.e., half the size of a full T layer. e) Topography of the cleaved ½T layer of 1T1H revealing the imperfect tetragonal lattice. The large white spot corresponds to the residue of the removed top ½T surface and dark spots correspond to vacancy sites. f) Topography of the H layer of 1T1H with hexagonal lattice after removing the unstable ½T layer by continuously sweeping the area with STM tip. The periodicity of the imaged T and H lattices is also visible in the Fast Fourier Transform of (e) and (f) along with more intricate details stemming from the misfit properties, described in Extended data Figs. 1 and 2. Scalebars in c), e) and f) represent 5 nm.

In 1T2H, the H-T-H (NbSe$_2$-LaSe-NbSe$_2$) trilayers are vertically stacked via van der Waals (vdW) interactions[15], whereas 1T1H forms a fully 3D ionocovalent crystal[18]. This is corroborated by our first-principles calculations of the cohesive energy. The calculated energy to split the isolated T layer into two ½T layers (Fig. 1d) is $E_T \approx -56.8$ meV/Å$^2$, which is lower than the cohesive energy $E_{T-H} \approx -100$ meV/Å$^2$ that binds the T and H layers. This prediction is consistent with our scanning tunnelling microscopy (STM), illustrated in Fig. 1 c-f, where the ½T layer is seen to be much less stable than the H surface layer. Indeed, we were unable to image the ½T layer unless the sample was cleaved at liquid nitrogen temperature and subsequently scanned in-situ in ultra-high vacuum.

The excess stoichiometry of the T layer (1.14) with respect to the H layer suggests a charge transfer of 0.57 $e^-$ per Nb, from the T layer into the two adjacent H layers in 1T2H, with an estimated carrier density $n_{est} = 4.6 \times 10^{21}$ cm$^{-3}$ based on the NbSe$_2$ unit cell lattice parameters[15]. The 1T2H Hall coefficient at 300 K is $R_H \approx + 1.45 \pm 0.1 \times 10^{-9}$ m$^3$/C (Fig. 2 a), corresponding to a charge carrier density $n \approx 4.7 \times 10^{21}$ cm$^{-3}$. The positive sign is consistent with transport occurring via the Nb *4d* hole pocket and the measured carrier density agrees with our estimate above, and prior measurements[14].

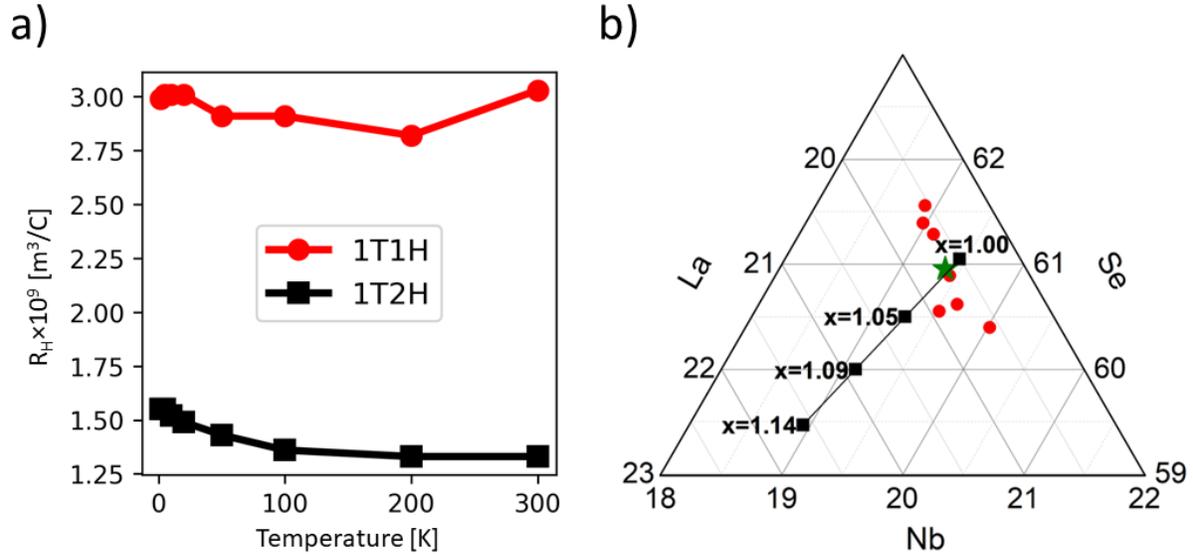

**Figure 2 – Hall measurements and stoichiometry.** a) Hall coefficients of 1T1H (red dots) and 1T2H (black squares). b) Projections of energy dispersive X-ray spectroscopic analyses for $La_xSe_{1.14}NbSe_2$ crystals in the La–Nb–Se ternary diagram (atomic percentages). Red dots are the mean analyses of each crystal while the green star represents the mean value for the seven crystals. The black line represents the theoretical compositions calculated between the vacancy-rich (x=1) and the vacancy-free (x=1.14) poles. The mean experimental composition lies on this line and evidences an amount of La vacancies around 11.1%. More details in Extended data Table 1.

For 1T1H, the same arguments suggest that each H layer accepts 1.14 $e^-$ per Nb from the two adjacent ½T layers. However, based on the formal oxidation states, each Nb can accept a maximum of 1 $e^-$, leaving 0.14 $e^-$ per La within the T layer. First-principles calculations confirm these considerations and show the Nb *4d* hole band at the *K*-point is occupied (Extended data Fig. 3a), while the partially filled La *5d* electron band crosses the Fermi level at the *Γ*-point (Extended data Fig. 3b). Hall measurements indicate mobile holes with a Hall coefficient $R_H \approx +3*10^{-9}$ m$^3$/C (Fig. 2 a), in contrast with the negative Hall coefficient expected if transport occurs via the LaSe conduction band. The carrier density at 300 K from the Hall coefficient is n $\approx 2*10^{21}$ cm$^{-3}$, or 0.75 $e^-$ per Nb atom. Though this is larger than the carrier density in the 1T2H structure, it is well below the maximum possible doping of 1 $e^-$ per Nb.

While evidence of La with a formal valence lower than 3 has been found in other layered compounds[19], a more plausible explanation of this discrepancy are deviations from stoichiometry. Point defects that act as acceptors can compensate the charge transferred into NbSe$_2$ and lower the effective doping. We consider the role of La vacancies in LaSe, motivated by the fact that prior work on other misfit compounds identified cation vacancies[20–22]. Figure 2b summarizes our EDS measurements of sample composition. Despite small variations over different crystals, the mean composition indicates a La vacancy concentration of 11.1%, suggesting that 0.76 electrons were transferred per Nb atom, in agreement with the Hall data.

Our ARPES measurements of 2H-NbSe$_2$, 1T2H and 1T1H (Fig. 3 a-c) show the Nb 4$d$ states crossing $E_F$ shift towards higher binding energies, as expected with electron doping of the H layer. Furthermore, the Nb 4$d$ hole band is not completely filled in the 1T1H structure, consistent with our positive Hall coefficient.

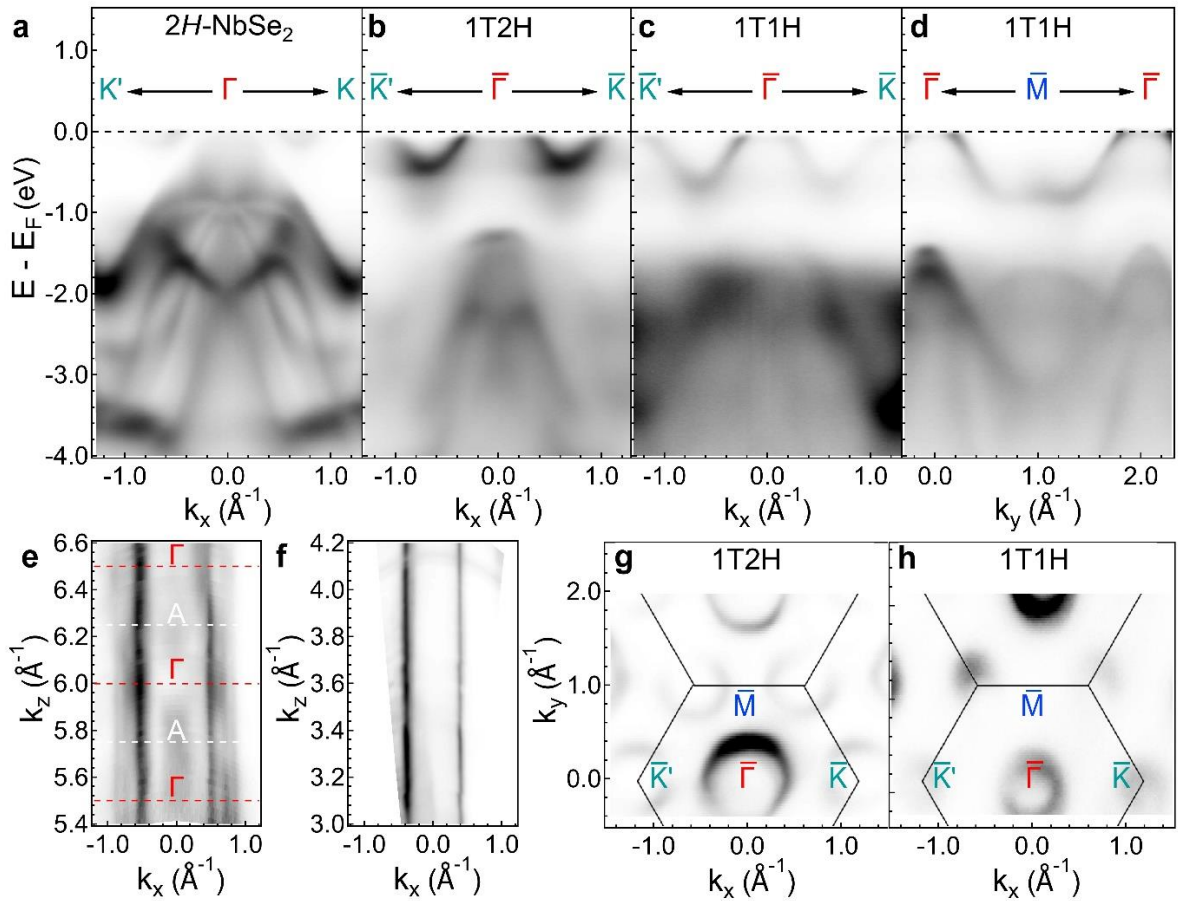

**Figure 3 – ARPES measurements of bulk NbSe$_2$ and misfit compounds.** (a) ARPES map along the $K' - \Gamma - K$ high-symmetry line of the hexagonal Brillouin zone of 2H-NbSe$_2$. h$\nu$ = 106 eV. (b) ARPES map along $\overline{K'} - \overline{\Gamma} - \overline{K}$ of 1T2H [(LaSe)$_{1.14}$(NbSe$_2$)$_2$]. (c-d) ARPES maps along $\overline{K'} - \overline{\Gamma} - \overline{K}$ and $\overline{\Gamma} - \overline{M} - \overline{\Gamma}$ of 1T1H [(LaSe)$_{1.14}$(NbSe$_2$)]. h$\nu$ = 150 eV for both misfits. (e-f) Out-of-plane momentum ($k_z$) dispersions at $k_y = 0$ Å$^{-1}$ and $E_F$ for NbSe$_2$ and 1T2H. h$\nu$ = 100-160 eV and h$\nu$ = 20-60 eV for 2H-NbSe$_2$ and 1T2H, respectively. An inner potential $\nu_0$ of 17.6 eV[23] and a work function value of 5.9 eV[24] were used for converting the photon energies into $k_z$ values within a free-electron final state approximation. The integer and half-integer values of $k_z$ correspond to the $\Gamma$ points in 2H-NbSe$_2$ (out-of-plane lattice parameter c = 12.5 Å)[25]. (g-h) Fermi surfaces of 1T2H (g) and 1T1H (h) with overlaid hexagonal surface Brillouin zones of 2H-NbSe$_2$ and associated $\overline{\Gamma}, \overline{K}, \overline{K'},$ and $\overline{M}$ high-symmetry points.

At the $\overline{K}(\overline{K'})$ points, $E_F$ lies very close to the top a Nb 4d band. In bulk 2H-NbSe$_2$ the Fermi surface around the $\overline{K}$ points is comprised of two spin-degenerate trigonally-warped barrels formed by the Nb 4d bands[26]. These split bands are visible around $k_x = \pm 0.8$ Å$^{-1}$ in the out-of-plane momentum ($k_z$) map. Fig. 3 e shows their weak, but finite, $k_z$-dispersion at the outer $\Gamma$ pocket around $k_x = \pm 0.5$ Å$^{-1}$. While this $k_z$ distortion is substantially smaller than the one found in the bulk calculations (see Extended data Fig. 3), one should keep in mind that the penetration depth in ARPES measurements is smaller than the interlayer distance and for this reason a perfect agreement is not expected. The inner $\Gamma$ pocket derived from the Se $4p_z$ orbital forms the highly three-dimensional "pancake" FS. The 2D nature of the electronic structure of the 1T2H misfit compound is evident in Fig. 3 f. Only the outer $\Gamma$ pocket remains, around a reduced $k_x$ value of ±0.4 Å$^{-1}$, due to the electron doping, and no $k_z$ dispersion is visible.

Using the Fermi wave-vectors, $k_F^{1T2H}$ and $k_F^{1T1H}$, of the misfit's $\overline{\Gamma}$ pockets in Figure 3 g, h we estimate the charge transfer in 1T1H by comparing the areas of the FS. We obtain a ratio of the hole densities $\frac{n_{2D}^{1T1H}}{n_{2D}^{1T2H}} = \left(\frac{k_F^{1T1H}}{k_F^{1T2H}}\right)^2 \approx 0.43$. Assuming that LaSe transfers 0.57 $e^-$ per Nb atom in 1T2H, the charge transfer obtained from ARPES in 1T1H is 0.82 $e^-$ per Nb atom, which compares well with the 0.75 $e^-$ obtained from the Hall and EDS measurements.

Figure 4 illustrates the calculated band structure of the monolayer NbSe$_2$, stoichiometric 1T2H and off-stoichiometric 1T1H with a La vacancy concentration of 12.5% comparable to the experimentally estimated value of 11.1%. For the misfit compounds, a large amount of electron doping into the H layer shifts the $E_F$ towards the top of the Nb 4d hole band at the K point, consistent with our ARPES and Hall measurements.

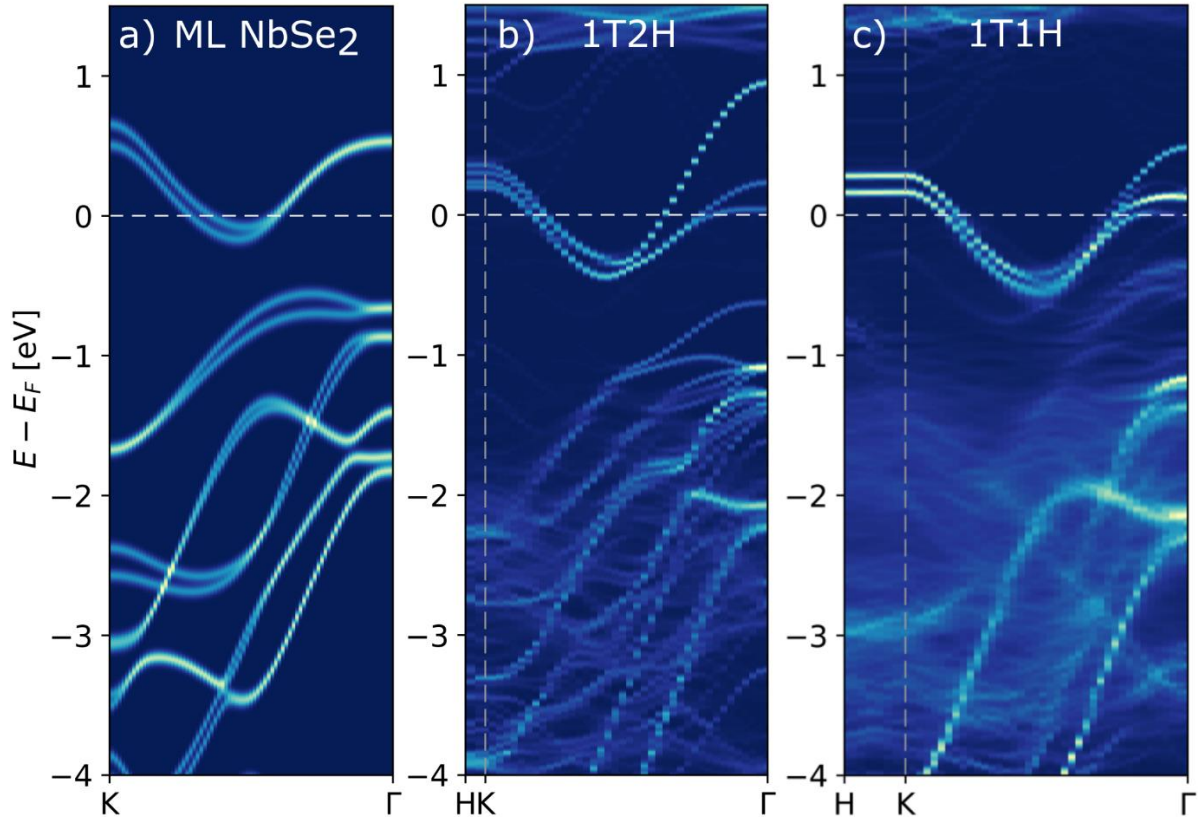

**Figure 4 – First-principles calculations.** Electronic structures unfolded onto the hexagonal Brillouin zone along high symmetry lines. a) Band structure for supercell calculations of bare monolayer of NbSe$_2$ b) 1T2H and c) 1T1H with 12.5% La vacancies.

## Discussion

To provide insight into how Ising protection arises in these bulk compounds we start with the 1T1H structure. Consider a NbSe$_2$ bilayer, where the combination of inversion symmetry and SOC leads to a pair of doubly degenerate states at $K$ and $K'$, and an interlayer hopping, which we approximate as a constant, $t$. The magnitude of $t$ in the 1T1H compound is different from that in the pristine NbSe$_2$ bilayer, since the separation between the H layers increases by ~ 5.6 Å and the interlayer tunnelling proceeds thorough the T layer rather than vacuum. We can use a 4x4 Hamiltonian to represent the spin up and down states in each NbSe$_2$ layer and calculate the electronic structure. By accounting for SOC, effective interlayer hopping and a Zeeman term due to an applied magnetic field one can show the Pauli susceptibility has two principal contributions: one from electrons at $E_F$, and the other from the states split-off by SOC that are farther away from $E_F$. The former term disappears in the superconducting state, the latter does not. The thermodynamic balance that determines the critical field is defined as $(\chi_n - \chi_s)B_t^2 = N(E_F)\Delta^2$ [27], where $\chi_n =$

$\chi_{Pauli}$ is the susceptibility in the normal state, $\chi_s$ is the susceptibility in the superconducting state and $B_t$ is the thermodynamic critical field. For a bilayer with SOC and an effective interlayer hopping $t$, we get $\chi_s = \chi_{Pauli} \frac{\zeta^2}{\zeta^2+t^2}$ and by averaging over $k_z$, for the bulk, we get $\chi_s = \chi_{Pauli} \frac{\zeta}{\sqrt{\zeta^2+t^2}}$. In 1T1H the spin-split bands at the $K$ point are separated by ~150 meV (Fig. 4 c), therefore $\zeta \sim 75$ meV. Now, $t$ is ¼ of the bandwidth along the out-of-plane direction. The out-of-plane bandwidth is $k$-dependent, the dispersion is flat along $K - H$, i.e., $t \sim 0$, while it is finite but small along $\Gamma - A$. We estimate the effective hopping relevant for the Ising protection as $t \sim 25$ meV. This leads to an enhancement of the critical field over $B_P$ by a factor $B_P \frac{\sqrt{\zeta^2+t^2}}{t} \sim 3B_P$. Therefore, the 1T1H bulk crystal can be considered as an "infinite" stack of rigidly doped NbSe$_2$ monolayers with extremely weak interlayer hopping via LaSe layers. This Ising protection due to reduced interlayer hopping is also consistent with the conjecture by Xi et al.[5] regarding the Pauli-violation for bilayer and trilayer NbSe$_2$.

In 1T2H, the structure can be viewed as a stack of centrosymmetric 2H-NbSe$_2$ bilayers separated by LaSe misfit layers (Fig. 1 b). Here we replace $t$ with a much larger number $t_0$ to account for the stronger interlayer coupling within the bilayer structure. If one assumes that $t_0 \sim 2\zeta$, as it is in bulk NbSe$_2$, we find the critical field is enhanced by a factor of $1.12 B_P$, lower than the factor of 5 observed in our experiments. To resolve this discrepancy, we account for the fact that the LaSe layers below and above the NbSe$_2$ bilayer are oriented differently with respect to NbSe$_2$, thus breaking the total inversion symmetry in the misfit[15]. We emulate this effect in our theory by applying a constant energy shift of $\pm T_0$ to the upper and lower layers of the bilayer, which lifts the Kramers' degeneracy. In Fig. 4b, two pairs of NbSe$_2$ bands are present at the $K$ point. Each pair is spin-split by ~ 110 meV and they are mutually shifted by ~ 50 meV at the $K$ point (Fig. 4 b).

This results in two effects: one, which is often considered in the discussion of single monolayers[28], leads to a canting of the spin directions away from the c-axis. The resulting spin can be projected onto the out-of-plane direction, yielding an Ising component, and onto the ab-plane, yielding planar components (Rashba and Dresselhaus). The latter have no Ising protection for in-plane magnetic fields[28], so it is often believed that breaking the intralayer mirror symmetry is detrimental for IS.

What has been missed previously is the second effect, which helps restore the Ising protection. If we solve our 4x4 Hamiltonian including the effect of $T_0$, one can show that the Ising protection is now determined by the larger of the two expressions; either $B_t = B_P \frac{\sqrt{\zeta^2+t_0^2}}{t_0}$ or $\mu_B B_t =$

$\sqrt{t_0^2 + (T_0 + \zeta^2)^2} - \sqrt{t_0^2 + (T_0 - \zeta^2)^2}$ (D.W., D. Agterberg, I.I.M, in preparation). In our case, $T_0$ ~ 25 meV at the *K* valley. However, since it is *k* dependent, we consider $T_0$ ~ 10 meV as a lower bound. This, combined with $\zeta \sim 55\ meV$ and $t_0 = 2\zeta$, yields $B_t$ ~ 150 T, demonstrating that a finite energy shift due to asymmetric encapsulation provides strong Ising protection. This Ising protection is likely to be further reduced due to an admixture of planar SOC[28] as we discuss above, but it is unlikely to overcome the large value of $B_t$.

Hence, this hierarchy of energy scales of interlayer coupling, namely, breaking of inversion symmetry by the LaSe layer, SOC and the magnitude of the superconducting gap in the 1T1H and 1T2H compounds provides a consistent picture for the strong Ising protection observed in these bulk compounds. This is different from the IS in the monolayer where only the last two energy scales are present. The presence of charge doping in these misfit compounds, larger than what can be achieved with traditional substitutional doping, introduces the possibility of shifting the Fermi level to be in-between the two spin-split bands, which is predicted to give rise to topological superconductivity[4,29,30]. It is plausible that these insights can be extended to other misfit compounds and bulk structures comprised of transition metal dichalcogenide layers where large in-plane critical magnetic fields that exceed the Pauli limit have been reported on but not identified as evidence of Ising superconductivity. This includes works on $(SnSe)_{1.16}(NbSe_2)_2$[31], $Ba_6Nb_{11}S_{28}$[32], cation intercalated $NbSe_2$[33] and organometallic intercalated compounds of 2H-$TaS_2$ [34,35]. Future experiments and theory work to determine the electronic and structural properties for this broad range of potential bulk Ising superconductors will be of great interest.

## Methods

### Sample preparation

Single crystals of $(LaSe)_{1.14}(NbSe_2)$ and $(LaSe)_{1.14}(NbSe_2)_2$ were grown by chemical vapor transport using iodine and the product of the solid-state reaction of the elemental precursors (i.e., La, Nb, Se). More details of the 1T2H sample preparation are described in ref.[14]. 1T1H was prepared identically except for the molar ratio which was changed to La/Nb/Se = 1.14/1/3.14.

### Energy-dispersive X-ray spectroscopy (EDS)

The chemical composition of 1T1H samples was inspected by energy-dispersive X-ray spectroscopy on a JEOL 5800LV scanning electron microscope equipped with a PGT microanalyzer and operating at 20 keV. Large crystals were pasted and mechanically pressed into carbon tape to reach a good horizontal alignment and improve accuracy.

### Scanning tunnelling microscopy (STM)

Scanning tunnelling microscopy data was acquired in ultra-high vacuum at a base pressure lower than $10^{-10}$ mbar and a base temperature of 1.14 K using a mechanically cut gold tip operated in the constant-current mode. Bias voltage of 1 V was applied to the sample, setpoint was 50 pA. The sample was mechanically cleaved in-situ at 77 K.

### Hall measurements

The magnetotransport measurements were realized in a Quantum Design Physical Properties Measurement System using AC transport and resistivity options in the temperature range from 1.8 K to 300 K. Magnetic fields up to 9 T were applied in various orientations.

### Angle-resolved photoemission spectroscopy (ARPES)

ARPES experiments were carried out at the CASSIOPEE beamline of the SOLEIL synchrotron radiation center, in Saint-Aubin, France. The electron detector was a Scienta R4000 hemispherical analyser with an angular acceptance of ± 15°. The experiments were performed at T = 14 K with a linear horizontal (LH) polarization of the incident light. The total energy resolutions were between 30 meV and 80 meV depending on the measurements.

**First-principles calculations**

The electronic structure was calculated by means of density functional theory as implemented in the Vienna ab-initio simulation package VASP.6.2[36]. The PAW pseudopotentials[36] were used with PBE exchange-correlation functional[37]. The lattice parameters and atomic positions within the structural models were relaxed with the force threshold below $2\times10^{-2}$ eV Å$^{-1}$. Dispersive van der Waals energy correction[38] was used for the 1T2H systems. The band structure unfolding was performed using the VASPKIT[39]. The cleavage energy calculations were obtained from total energy calculations for a series of finite-size slabs with particular surface terminations.

## Acknowledgments

We thank V. Tkáč and J. Cordiez for assistance with the Hall measurements and chemical analysis, respectively. This work was supported by the projects APVV-20-0425, VEGA 2/0058/20, EMP-H2020 Project No. 824109, COST action CA21144 (SUPERQUMAP), EU ERDF (European regional development fund) Grant No. VA SR ITMS2014+ 313011W856. D.W. was supported by the Office of Naval Research (ONR) through the Naval Research Laboratory's Basic Research Program. I.I.M. was supported by ONR through grant N00014-20-1-2345. C.M. and G.K. acknowledges financial support from the Swiss National Science Foundation (SNSF) Grant No. P00P2_170597. M.G. and J.H. acknowledge support by the projects APVV-SK-CZ-RD-21-0114, VEGA 1/0105/20 and Slovak Academy of Sciences project IMPULZ IM-2021-42 and project FLAG ERA JTC 2021 2DSOTECH. S.S, L. C. and T. C. acknowledge financial support from the Agence Nationale de la Recherche (ANR) Project No. ANR-21-CE30-0054-02.

## Author contributions

T.S., P.S. conceived the study, D.V., S.S., L.C. synthesized the samples, carried out and analysed EDS experiments, O.Š. performed and analysed the STM experiments, M.K. performed Hall measurements, T.S. supervised and interpreted the STM experiments and Hall measurements, T.J., G.K., T.C. performed the ARPES experiment with the help of P.LeF. and F.B., T.J., C.M., G.K. analysed and interpreted the ARPES data, M.G., J.H., M.C. performed the first-principles calculations, P.S., P.Sz., T.C. designed the scientific objectives and oversaw the experiments, D.W., I.I.M developed the theoretical model, T.S., D.W. and I.I.M. wrote the manuscript with input from all co-authors.

| Crystal (n) | Atomic composition (%) | | | Chemical formula | | |
|---|---|---|---|---|---|---|
| | Se | Nb | La | Se | Nb | La |
| Crystal 1 (5) | 61,29(12) | 19,11(7) | 19,60(7) | 3,140(-) | 0,979(6) | 1,004(6) |
| Crystal 2 (5) | 60,6(2) | 19,52(14) | 19,92(10) | 3,140(-) | 1,012(10) | 1,033(8) |
| Crystal 3 (7) | 61,56(5) | 18,90(6) | 19,54(7) | 3,140(-) | 0,964(4) | 0,997(4) |
| Crystal 4 (5) | 61,39(12) | 18,97(13) | 19,64(4) | 3,140(-) | 0,970(9) | 1,005(4) |
| Crystal 5 (7) | 60,62(13) | 19,64(9) | 19,74(6) | 3,140(-) | 1,017(7) | 1,022(5) |
| Crystal 6 (8) | 60,89(13) | 19,44(9) | 19,67(10) | 3,140(-) | 1,002(7) | 1,014(7) |
| Crystal 7 (5) | 60,4(2) | 20,0(2) | 19,58(7) | 3,140(-) | 1,041(12) | 1,018(7) |
| **Average** | **60,96(13)** | **19,37(10)** | **19,67(7)** | **3,140(-)** | **0,998(7)** | **1,013(6)** |
| **11,1% of La vacancies** | **60,93** | **19,40** | **19,67** | **3,140** | **1** | **1,014** |

**Extended data Table 1 – Chemical composition of the 1T1H samples determined by EDS.** Atomic percentages with relevant errors in brackets obtained by EDS analysis for seven different crystals of 1T1H ($n$ is the number of measurements on each single crystal). To calculate the chemical formula, the amount of Selenium was fixed to 3.14 according to the theoretical chemical composition $(LaSe)_{1.14}(NbSe_2)$. The last row represents the theoretical values of the atomic percentage and chemical formula for the case of a crystal with 11.1% La vacancies.

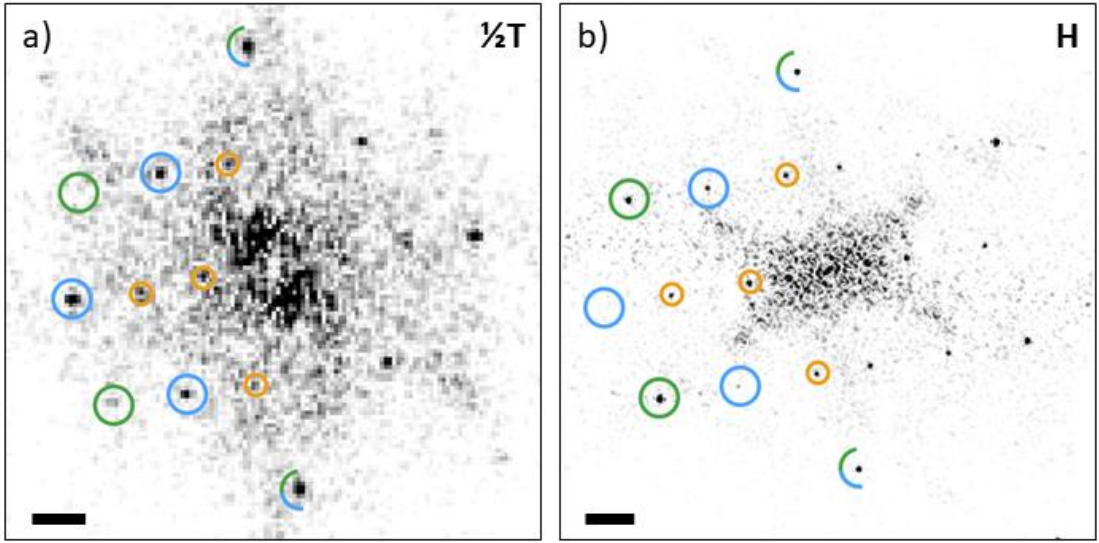

**Extended data Fig. 1 – a) and b) show the Fourier transform modulus of the topography in Fig. 1 e) and f), respectively.** In both FFT images both hexagonal and square Bragg peaks are present, denoted by green and blue circles, respectively, as well as peaks corresponding to the lattice modulations due to misfit, marked by orange circles. The blue-green semicircles belong simultaneously to the square and hexagonal lattice and lie in the commensurate b direction. The fact that both square and hexagonal lattices are present in both FFT images demonstrates the surface lattice modulation induced by the underlying layer. The scalebar represents 5 nm$^{-1}$.

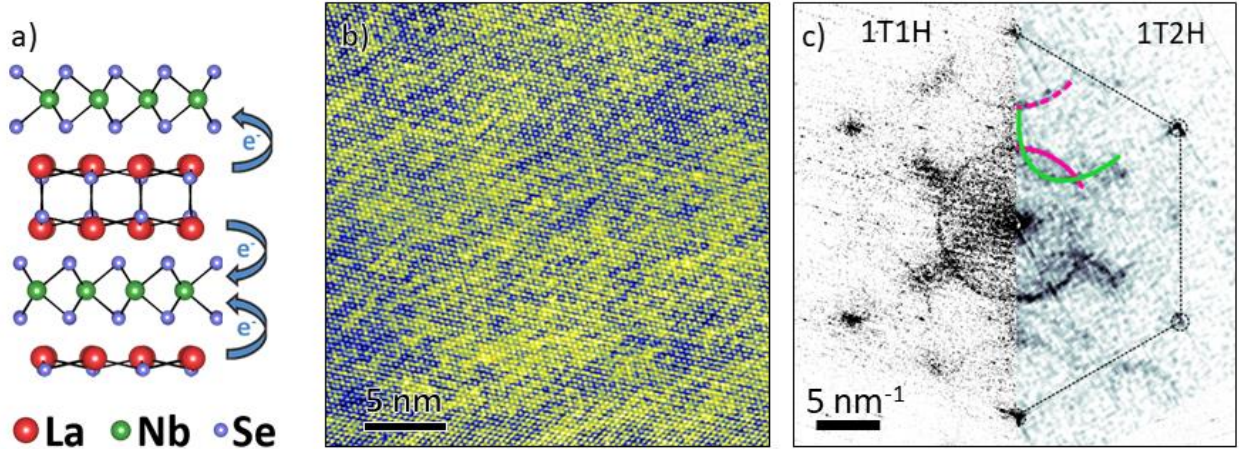

**Extended data Fig. 2 – a) Crystal structure of 1T1H without the topmost ½T layer. b) STM topography of the H layer of 1T1H after the unstable ½T layer was removed by continuously sweeping the area with STM tip (30nm x 30nm, $U_{bias}$ = 0.5 mV, $I_{set}$ = 80 pA, T = 77 K). c) left: Fourier transform modulus of b); right: Fourier transform modulus of 1T2H topography from ref.[14].** When the topmost ½T layer of 1T1H is removed, the H surface is charge-doped only by the underlying ½T layer, thus mimicking the situation in 1T2H as demonstrated in a). Because the topography in b) was acquired at low bias, i.e., energy close to $E_F$, it is possible to resolve the quasiparticle interference patterns in its Fourier transform. The fact that this pattern is almost identical to the 1T2H case, as demonstrated in c), is well in line with the notion of the half-doped H surface layer. Though this is only a local surface effect not affecting the bulk properties of the crystal, it validates our qualitative understanding of the charge transfer between layers.

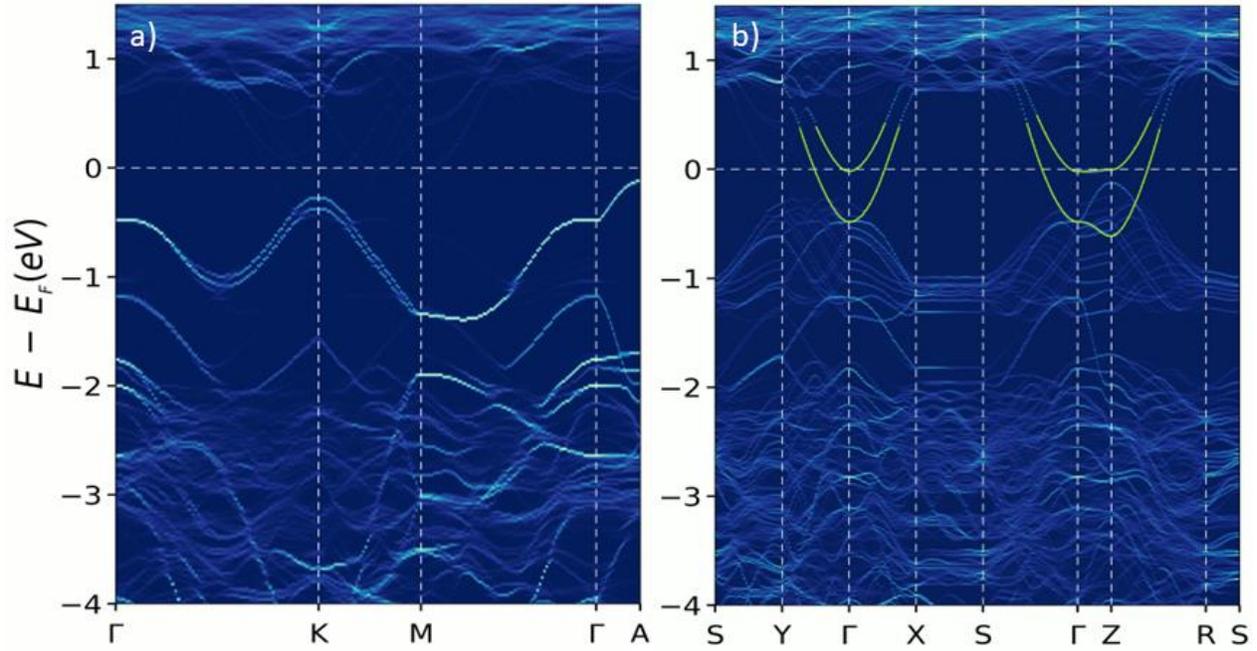

**Extended data Fig. 3 – Electronic band structure along high symmetry lines for stoichiometric bulk 1T1H calculated from first principles.** Band structures are unfolded to a) hexagonal and b) orthorhombic Brillouin zones corresponding to zones for NbSe$_2$ and LaSe crystals. The partially filled La electron bands crossing the Fermi level at the $\Gamma$-point are highlighted in green.